\documentclass[prx,longbibliography,showpacs,onecolumn, floatfix,superscriptaddress,aps,11pt]{revtex4-2}

\usepackage[utf8]{inputenc}
\usepackage{color}
\usepackage{amssymb,amsmath,mathtools}
\usepackage{overpic}
\usepackage{booktabs, array, tabularx, lipsum,multirow}
\usepackage{dcolumn}
\usepackage{epstopdf}
\usepackage{bm}
\usepackage{ulem}
\usepackage{graphicx}
\usepackage{threeparttable}
\usepackage{xcolor}
\usepackage{makerobust}
\usepackage[caption=false]{subfig}
\usepackage{hyperref}
\usepackage{dblfloatfix}
\usepackage{xr}
\usepackage{comment}
\newcommand{\captiontitle}[1]{\textbf{#1}}

\newcommand{\makeauthor}[2]{\newcommand{#1}[1]{{%
  \protect%
  \color{#2}{%
    \bfseries
  } ##1}}%
  \MakeRobustCommand#1}
\makeauthor{\af}{orange}
\makeauthor{\qlxu}{black}
\usepackage{float}

\newcommand{\icon}[2][1.10cm]{{\includegraphics[width=#1]{#2}}}

\begin{document}

\title{Organizing Principles for Moiré Quantum Matter}

\author{Qiaoling Xu}
\altaffiliation{These authors contributed equally.}
\affiliation{Tsientang Institute for Advanced Study, Zhejiang 310024, China}
\affiliation{College of Physics and Electronic Engineering, Key Laboratory of Micro-Nano Optoelectronic Materials and Devices, Center for Computational Sciences, Sichuan Normal University, Chengdu 610068, China}

\author{Yifan Gao}
\altaffiliation{These authors contributed equally.}
\affiliation{Max Planck Institute for the Structure and Dynamics of Matter, Luruper Chaussee 149, 22761 Hamburg, Germany}
\affiliation{Tsientang Institute for Advanced Study, Zhejiang 310024, China}

\author{Tao Zhang}
\altaffiliation{These authors contributed equally.}
\affiliation{Department of Materials Science and Engineering, City University of Hong Kong, Kowloon Tong, Hong Kong, SAR 999077, P. R. China}
\affiliation{Songshan Lake Materials Laboratory, 523808 Dongguan, Guangdong, China}

\author{Ammon Fischer}
\altaffiliation{These authors contributed equally.}
\affiliation{Max Planck Institute for the Structure and Dynamics of Matter, Luruper Chaussee 149, 22761 Hamburg, Germany}
\affiliation{Center for Computational Quantum Physics (CCQ), The Flatiron Institute, New York, New York 10010, USA}

\author{Yi Jiang}
\affiliation{Donostia International Physics Center (DIPC), Paseo Manuel de Lardizábal. 20018, San Sebastián, Spain}

\author{Hanqi Pi}
\affiliation{Donostia International Physics Center (DIPC), Paseo Manuel de Lardizábal. 20018, San Sebastián, Spain}

\author{Zike Fan}
\affiliation{Tsientang Institute for Advanced Study, Zhejiang 310024, China}
\affiliation{School of Physics, Liaoning University, Shenyang 110036, PR China}

\author{Dongdong An}
\affiliation{National Laboratory of Solid-State Microstructures, School of Physics, Nanjing University, Nanjing 210093, China}

\author{Kun Zhou}
\affiliation{Department of Chemical and Biological Engineering, The Hong Kong University of Science and Technology, Clear Water Bay, Hong Kong SAR, China.}
\affiliation{State Key Laboratory of Integrated Optoelectronics, Key Laboratory of Automobile Materials of MOE, College of Materials Science and Engineering, Jilin University, Changchun 130012, China}

\author{Yingjian Li}
\affiliation{State Key Laboratory of Integrated Optoelectronics, Key Laboratory of Automobile Materials of MOE, College of Materials Science and Engineering, Jilin University, Changchun 130012, China}

\author{Yongqing Li}
\affiliation{School of Physics, Liaoning University, Shenyang 110036, PR China}

\author{Yuhao Fu}
\affiliation{State Key Laboratory of High Pressure and Superhard Materials,
Key Laboratory of Material Simulation Methods \& Software of Ministry of Education, College of Physics, Jilin University, Changchun, 130012, China}

\author{Lei Wang}
\affiliation{National Laboratory of Solid- State Microstructures, School of Physics, Nanjing University, Nanjing 210093, China}

\author{Lijun Zhang}
\affiliation{State Key Laboratory of Integrated Optoelectronics, Key Laboratory of Automobile Materials of MOE, College of Materials Science and Engineering, Jilin University, Changchun 130012, China}

\author{B.~Andrei Bernevig}
\email[Email: ]{bernevig@princeton.edu}
\affiliation{Department of Physics, Princeton University, Princeton, NJ 08544, USA}
\affiliation{Donostia International Physics Center (DIPC), Paseo Manuel de Lardizábal. 20018, San Sebastián, Spain}
\affiliation{IKERBASQUE, Basque Foundation for Science, Bilbao, Spain}

\author{Dante M. Kennes}
\email[Email: ]{Dante.Kennes@rwth-aachen.de}
\affiliation{Institut f\"ur Theorie der Statistischen Physik, RWTH Aachen University and JARA-Fundamentals of Future Information Technology, 52056 Aachen, Germany}
\affiliation{Max Planck Institute for the Structure and Dynamics of Matter, Luruper Chaussee 149, 22761 Hamburg, Germany}

\author{Enge Wang}
\affiliation{Tsientang Institute for Advanced Study, Zhejiang 310024, China}
\affiliation{International Center for Quantum Materials, Collaborative Innovation Center of Quantum Matter, Peking University, Beijing, China}

\author{Angel Rubio}
\email[Email: ]{Angel.Rubio@mpsd.mpg.de}
\affiliation{Max Planck Institute for the Structure and Dynamics of Matter, Luruper Chaussee 149, 22761 Hamburg, Germany}
\affiliation{Initiative for Computational Catalysis, Simons Foundation Flatiron Institute, New York, New York 10010, USA}
\affiliation{Center for Computational Quantum Physics (CCQ), The Flatiron Institute, New York, New York 10010, USA}
\affiliation{Nano-Bio Spectroscopy Group and ETSF, Universidad del País Vasco UPV/EHU, San Sebastián 20018, Spain}

\author{Lede Xian}
\email[Email: ]{ldxian@tias.ac.cn}
\affiliation{Tsientang Institute for Advanced Study, Zhejiang 310024, China}
\affiliation{Max Planck Institute for the Structure and Dynamics of Matter, Luruper Chaussee 149, 22761 Hamburg, Germany}
\affiliation{Songshan Lake Materials Laboratory, 523808 Dongguan, Guangdong, China}
 
\date{\today}

\begin{abstract}

Moir\'e flat bands in van der Waals bilayers are usually discussed through a small set of mechanisms associated with the $\Gamma$ and $K$ valleys of hexagonal crystals, and more recently with $M$-valleys systems. Here we show that this view is incomplete. The momentum-space location and effective local orbital character of the monolayer's band edge, in conjunction with the moir\'e symmetry and the symmetry representations of the resulting bands, provide a general set of organizing variables for the emergent low-energy moir\'e Hamiltonian. Applying fully relaxed first-principles calculations, band unfolding and symmetry-representation analysis to more than 600 commensurate twisted bilayers spanning all 2D lattice classes, we identify several routes to moir\'e quantum matter beyond the conventional single-orbital paradigm. The resulting flat bands realize trigonal, honeycomb, square, checkerboard and kagome-like Hubbard models with single-orbital, multi-orbital and multi-site Hilbert spaces; spin-orbit-coupled multi-orbital flat bands exhibit symmetry-indicated topology beyond the conventional $K$-valley setting; and nonsymmorphic moir\'e symmetries enforce semimetallic flat-band connectivity. Analogous quasi-one-dimensional flat-band structures are found in $M$-valley hexagonal systems and $X$-valley square or rectangular systems resulting from emergent momentum-space nonsymmorphic symmetries. Separately, coupled multi-valley manifolds with kagome-like connectivity are identified in several systems whose parent band edges lie at non-high-symmetry points. These results establish a valley-orbital-symmetry framework for connecting parent-material electronic structure to emergent moir\'e Hamiltonians relevant to correlated, topological and symmetry-enforced moir\'e phases.

\end{abstract}

\maketitle

\section{\textbf{INTRODUCTION}}

Twisted van der Waals heterostructures have emerged as a versatile platform for engineering interacting quantum matter. Long-wavelength moir\'e superlattices suppress kinetic energy, restructure the single-particle Hilbert space and promote correlated and topological phases that are absent in the underlying monolayers. This capability has already enabled correlated insulators, superconductivity, Chern bands, heavy fermions, generalized Wigner crystals and one-dimensional correlated states in moir\'e systems \cite{cao2018correlated,cao2018unconventional,tang2020simulation,regan2020mott,li2021quantum,wang2020correlated,park2023observation,xu2023observation,zhao2023gate,wang2022one,zhao2024realization,wei2025valley,xia2026bandwidth,guo2026angle}. At the model level, moir\'e materials have been used to emulate honeycomb, triangular and square Hubbard models, multi-orbital bands, quasi-one-dimensional dispersions and spin-orbit-coupled topological bands \cite{xian2021realization,angeli2021gamma,ma2025relativistic,wu2018hubbard,tang2020simulation,xu2025engineering,kariyado2025single,kariyado2026moiregammavalleysquarelattice,bao2026moir,kennes2020one,claassen2022ultra,gao2022,cualuguaru2025moire,kennes_moire_2021}. Complementary high-throughput, database and machine-learning efforts have begun to explore this broader materials space \cite{bao2024deep,nakatsuji2025high,kaplan2025machine}, highlighting the need for a physics classification that connects narrow bands to the degrees of freedom and symmetries they carry.

The central open problem is how to organize this diversity. Much of the current intuition comes from a few material families and from $\Gamma$- or $K$-valley mechanisms in hexagonal crystals. These examples are crucial, but they do not by themselves answer which microscopic degrees of freedom survive in a general moir\'e miniband, where the corresponding Wannier orbitals are centered, or when symmetry forces topology or band connectivity. In other words, the key challenge is not simply to identify narrow bands, but to connect the valley momentum  and orbital content of the parent band edge with the symmetry, effective degrees of freedom and classification of the resulting moir\'e manifold.

We use the following organizing principle throughout the paper:

\begin{equation}
(\mathbf{k}_0, \mathcal O_{\mathbf{k}_0}, G_{\mathrm M}, \{\rho_{\mathbf{k}}\}_{\mathbf{k}\in\mathrm{HSP}_{\mathrm M}})
\longmapsto
(\mathcal W, \mathcal C, \mathcal{I}; \text{$\mathbf Q$-lattice}),
\label{eq:design-map}
\end{equation}

where $\mathbf{k}_0$ locates the parent band edge in momentum space, $\mathcal O_{\mathbf{k}_0}$ records the effective local orbital of the corresponding parent-band subspace, and $G_{\mathrm M}$ is the moir\'e space group. The remaining input collects the little-group irreducible representations (irreps) carried by the target moir\'e band manifold at the high-symmetry points (HSPs) of the moir\'e Brillouin zone. These quantities enter two complementary branches of the map. In the symmetry-classification branch, $G_{\mathrm M}$ and the calculated irreps are used to establish the Wannier characterization $\mathcal W$, band-connectivity classification $\mathcal C$ and topological classification $\mathcal I$. In the continuum-model branch, $\mathbf{k}_0$ fixes the principal geometry of the $\mathbf Q$-lattice, $\mathcal O_{\mathbf{k}_0}$ supplies its internal degrees of freedom, and $G_{\mathrm M}$ restricts the symmetry-allowed couplings. The semicolon distinguishes the symmetry-classification and continuum-model outputs, which provide complementary real-space and momentum-space descriptions of the low-energy manifold. Figure~\ref{Fig1}a provides a graphical realization of this map, showing how the relevant descriptors enter through the parent crystal, twisting and moir\'e reconstruction.

We implement this design map by combining density-functional theory (DFT), band unfolding and elementary band-representation (EBR) analysis\cite{Bradlyn2017TQC,Vergniory2017GraphTQC,Elcoro2017DoubleGroups}. We first identify the real-space lattice and Wannier characterization of the moir\'e flat-band manifolds and determine their symmetry-based connectivity and topology. We then turn to the momentum-space structures generated by boundary valleys and to the coupled multi-valley manifolds derived from non-high-symmetry-point (nHSP) band edges. This organization emphasizes distinct physical mechanisms rather than a material-by-material survey.

\section{\textbf{Extracting parent-valley design variables from first principles}}

\begin{figure*}[ht]
\includegraphics[width=\textwidth]{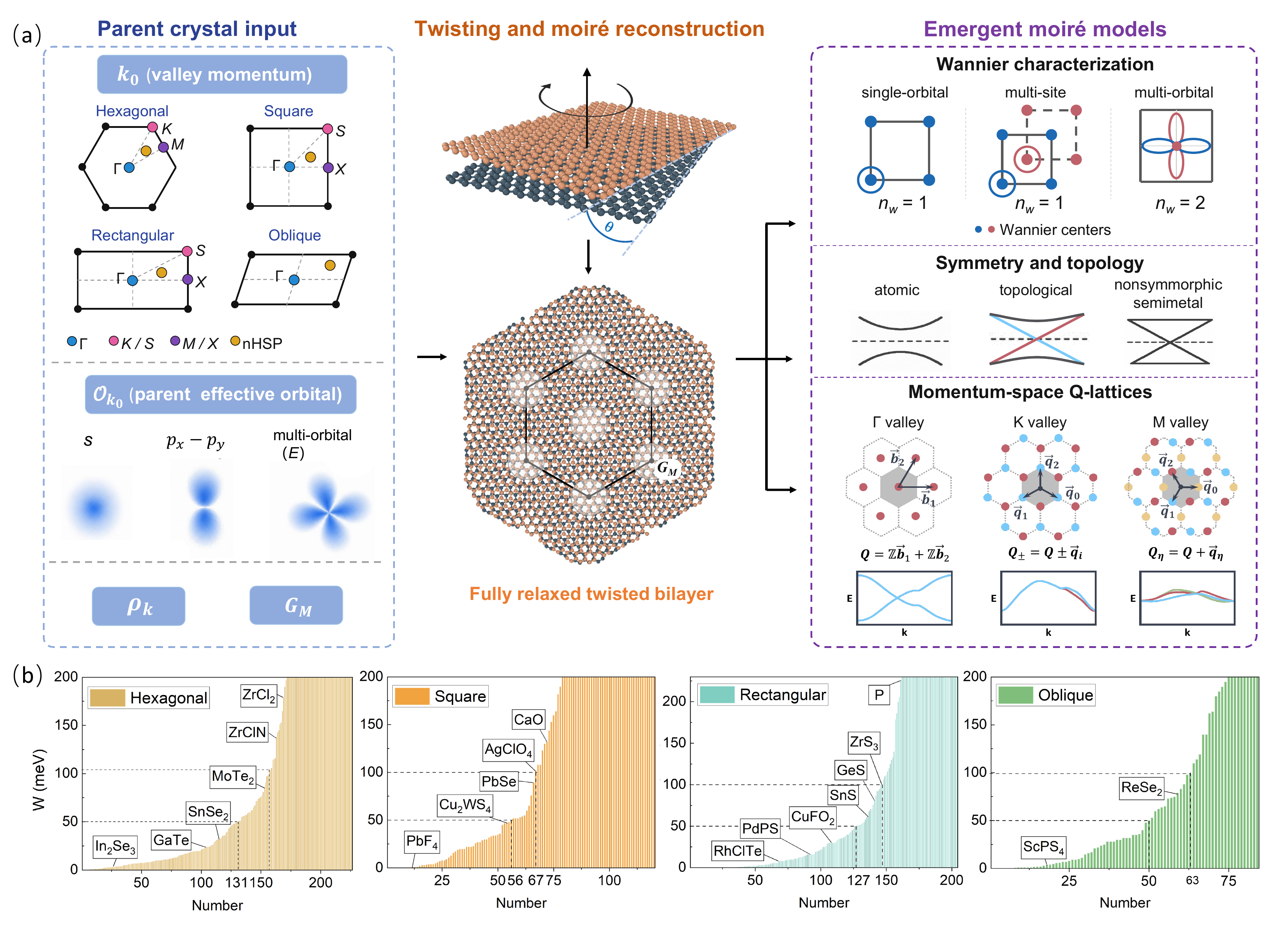}
\centering
\caption{ \fontsize{10pt}{11pt}\selectfont \captiontitle{Design map and first-principles atlas of emergent moir\'e models.} (a) Schematic representation of the organizing map in Eq.~\eqref{eq:design-map}. The parent band edge is specified by its valley momentum $\mathbf{k}_0$ and orbital content $\mathcal O_{\mathbf{k}_0}$, illustrated for hexagonal, square, rectangular and oblique crystals and for representative single- and multi-orbital states. Twisting and structural relaxation produce a commensurate moir\'e bilayer with space group $G_{\mathrm M}$, while the calculated minibands provide the little-group representations $\rho_{\mathbf{k}}$ at the high-symmetry momenta of the moir\'e Brillouin zone. These descriptors organize complementary characterizations of the emergent low-energy manifold: its Wannier content $\mathcal W$, including the number and positions of the Wannier centers; its band connectivity $\mathcal C$ and topological classification $\mathcal I$; and its valley-controlled momentum-space $\mathbf Q$-lattice. Representative outcomes include single-orbital, multi-site and multi-orbital Wannier models, atomic, topological and nonsymmorphic-semimetallic band structures, and the distinct $\mathbf Q$-lattices generated by $\Gamma$-, $K$- and $M$-valley band edges. (b) Band-edge bandwidth distributions for the four lattice classes. The systems are ordered by increasing bandwidth $W$ within each class. Horizontal dashed lines mark $W=50$ and $100$~meV, while the vertical dotted lines indicate the corresponding numbers of systems below these thresholds. Selected compounds are labeled to provide representative reference points across the different lattice classes and bandwidth ranges.}
\label{Fig1}
\end{figure*}

Guided by the design map summarized in Fig.~\ref{Fig1}a, we constructed a first-principles atlas of fully relaxed commensurate twisted bilayers. The starting point is a set of semiconducting monolayers selected from existing two-dimensional materials databases, including 2DMatPedia and MC2D \cite{zhou20192dmatpedia,mounet2018two,jiang24112d}. We retained materials whose corresponding three-dimensional bulk compounds have exfoliation energies below 164 meV/atom, approximately 1.5 times that of black phosphorus, and whose monolayers have finite gaps. These choices focus the analysis on band-edge moir\'e states that can be separated from metallic band entanglement and are, in principle, accessible by electrostatic gating.

The selected monolayers were grouped into hexagonal, square, rectangular and oblique lattice classes. For each class, commensurate moir\'e superlattices were generated using a generalized lattice-transformation construction, followed by full structural relaxation and DFT calculations with van der Waals corrections. The integer-matrix construction, commensurability criteria, relaxation protocol and computational details are given in the Supplementary Information (SI). Once a bilayer is relaxed, band unfolding traces the relevant minibands back to their parent valleys, while orbital projections are used where needed to characterize the parent states. Symmetry eigenvalues then supply the irrep data used in the EBR analysis. In this way, each calculation is converted into a low-energy-model assignment rather than merely a ranked materials candidate.

The resulting atlas contains more than 600 twisted bilayers, comprising 247 hexagonal, 121 square, 204 rectangular and 84 oblique moir\'e superlattices. The corresponding structural and electronic-structure data are compiled in the Twisted Bilayer Moiré Superlattice Database (TBMSD; https://2dstack.tias.ac.cn/tbmsd/). Figure~\ref{Fig1}b summarizes the band-edge bandwidth distributions. Although isolated flat bands do not occur in every material at the twist angles considered, more than 75\% of the surveyed systems exhibit band-edge bandwidths below 200 meV. This threshold defines a broad pool of narrow-band candidates. Rather than treating this bandwidth threshold as the final classification, we use the atlas to determine which low-energy lattice and moir\'e orbital models are realized, when symmetry enforces band connectivity or topology, and how boundary or nHSP valleys reshape the momentum-space structure of the problem.

\section{\textbf{Orbital and Wyckoff-position engineering of moir\'e Hubbard models}}

\begin{table}[ht]
\renewcommand{\arraystretch}{0.8}
\setlength{\tabcolsep}{3pt}
{\fontsize{9.5pt}{12.5pt}\selectfont
\centering
\begin{tabular}{cccccccc}
\toprule
\multirow{2}{*}{\textbf{Lattice}} &
\multicolumn{2}{c}{\multirow{2}{*}{\begin{tabular}[c]{@{}c@{}}\textbf{Number of Wyckoff pos.}\\\textbf{(representative models)}\end{tabular}}} &
\multirow{2}{*}{$n_w$} &
\multicolumn{4}{c}{\textbf{Representative materials}} \\
\cmidrule(lr){5-8}
& \multicolumn{2}{c}{} &&
\textbf{System ($G_{\mathrm M}$)} &
\textbf{Angle ($^\circ$)} &
\textbf{Band set pos.} &
\textbf{(E)BRs} \\
\midrule
\multirow[c]{10}{*}{Hexagonal}
& \multirow[c]{4}{*}{1 (trigonal)}
& \multirow[c]{4}{*}{\icon{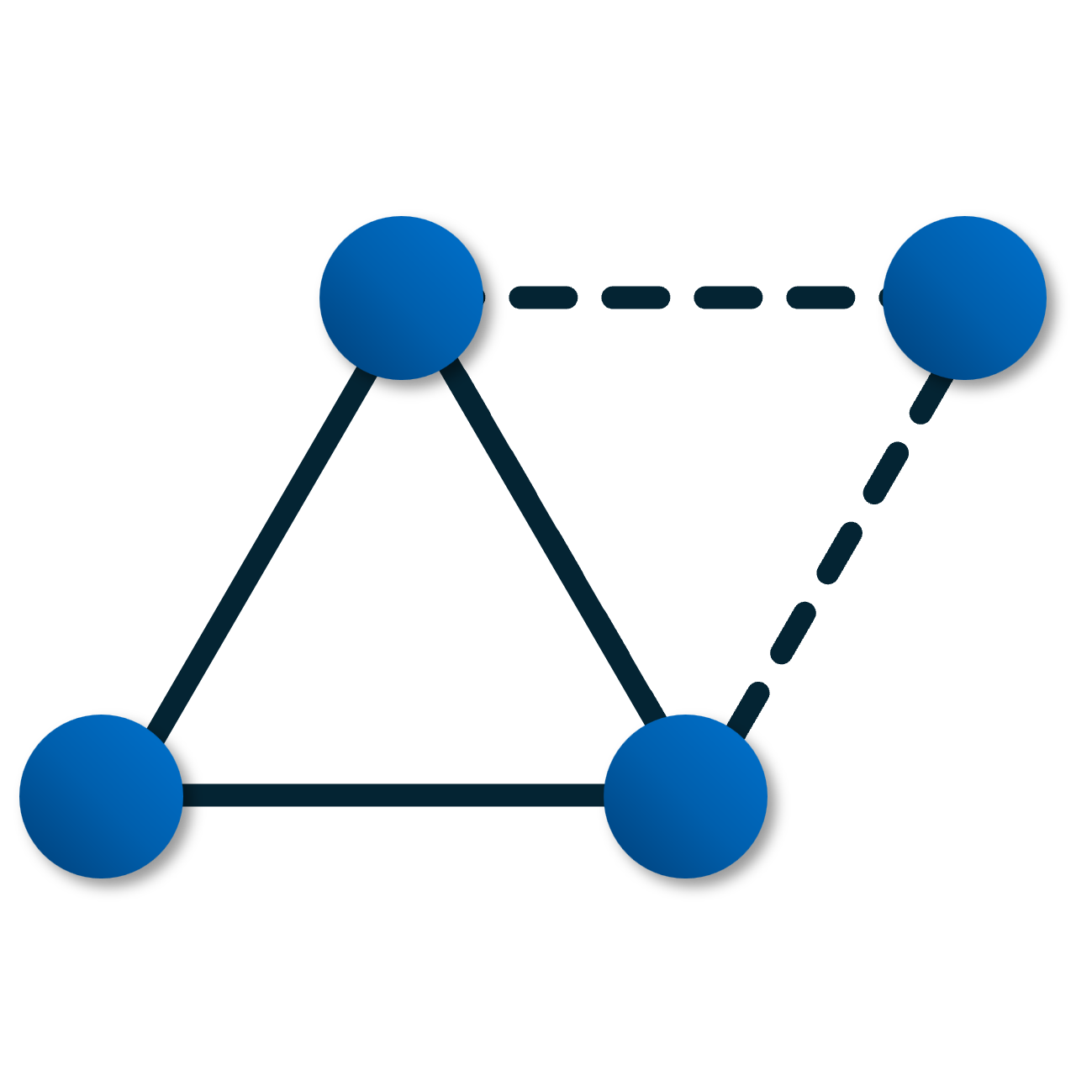}}
& \multirow[c]{2}{*}{1}
& PtSe$_2$
& \multirow[c]{2}{*}{7.34}
& \multirow[c]{2}{*}{VBM}
& \multirow[c]{2}{*}{$\bar{E}_1@1a$} \\
&&&& (P312) &&& \\

&&&
\multirow[c]{2}{*}{2}
& GaS(R)
& \multirow[c]{2}{*}{7.34}
& \multirow[c]{2}{*}{VBM-2}
& \multirow[c]{2}{*}{$\prescript{1}{}{\bar{E}}\prescript{2}{}{\bar{E}}@1a$, $\bar{E}_1@1a$} \\
&&&& (P312) &&& \\

\cmidrule(lr){2-8}

& \multirow[c]{4}{*}{2 (honeycomb)}
& \multirow[c]{4}{*}{\icon{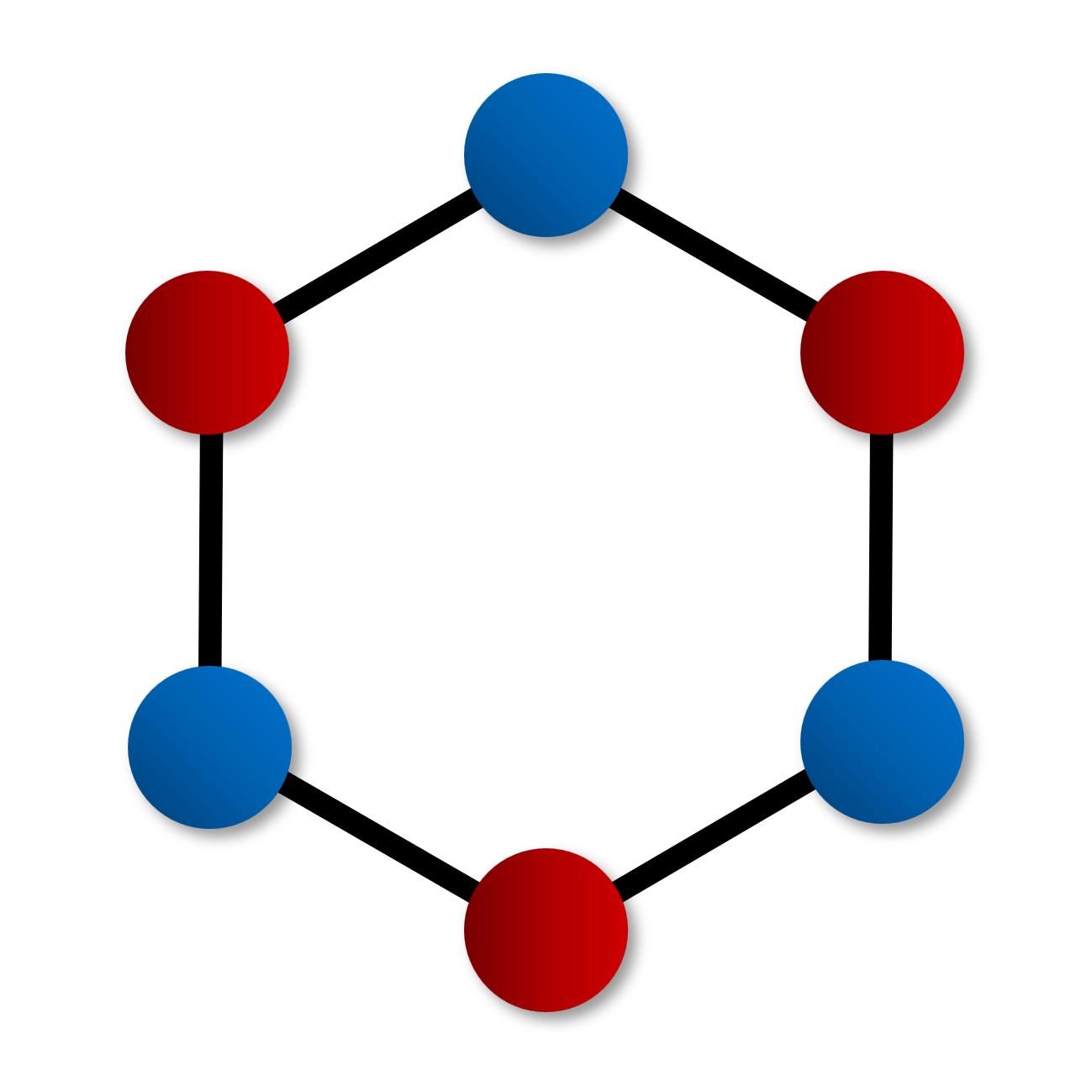}}
& \multirow[c]{2}{*}{1}
& WS$_2$
& \multirow[c]{2}{*}{7.34}
& \multirow[c]{2}{*}{VBM}
& \multirow[c]{2}{*}{$\prescript{1}{}{\bar{E}}\prescript{2}{}{\bar{E}}@2d$} \\
&&&& (P321) &&& \\

&&&
\multirow[c]{2}{*}{2}
& ZnO
& \multirow[c]{2}{*}{7.34}
& \multirow[c]{2}{*}{VBM}
& $\bar{E}\bar{E}@1a$, $\prescript{1}{}{\bar{E}}\prescript{2}{}{\bar{E}}@1a$ \\
&&&& (P3) &&&
$\bar{E}\bar{E}@1b$, $\prescript{1}{}{\bar{E}}\prescript{2}{}{\bar{E}}@1b$ \\

\cmidrule(lr){2-8}

& \multirow[c]{2}{*}{3 (Kagome)}
& \multirow[c]{2}{*}{\icon{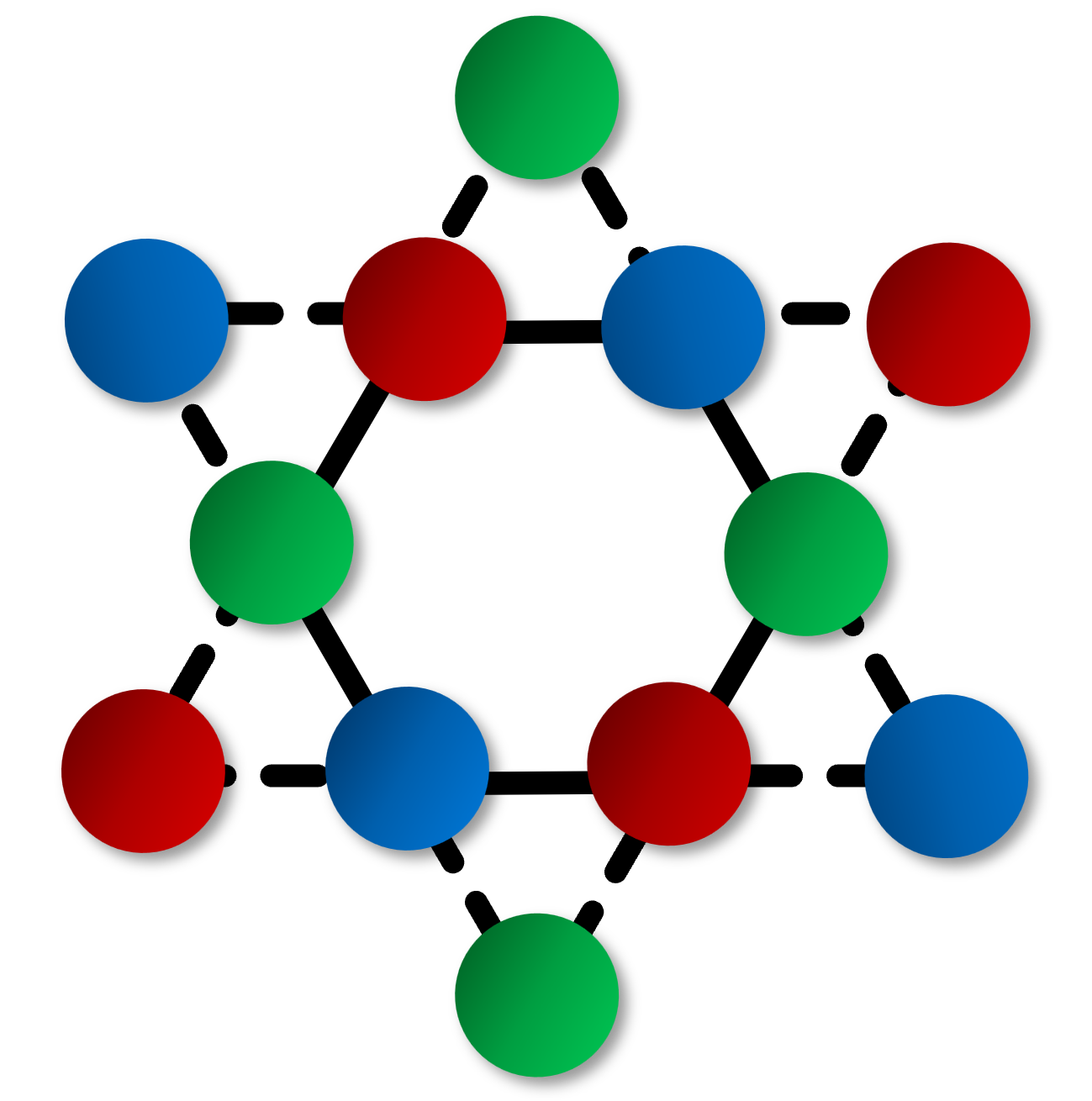}}
& \multirow[c]{2}{*}{1}
& TlSbO$_3$
& \multirow[c]{2}{*}{7.34}
& \multirow[c]{2}{*}{VBM}
& $\prescript{1}{}{\bar{E}}\prescript{2}{}{\bar{E}}@1a$, $2 \bar{E}_1@1a$ \\
&&&& (P321) &&&
or $\prescript{1}{}{\bar{E}}\prescript{2}{}{\bar{E}}@3e$ \\

\midrule

\multirow[c]{8}{*}{Square}
& \multirow[c]{6}{*}{1 (square)}
& \multirow[c]{6}{*}{\icon{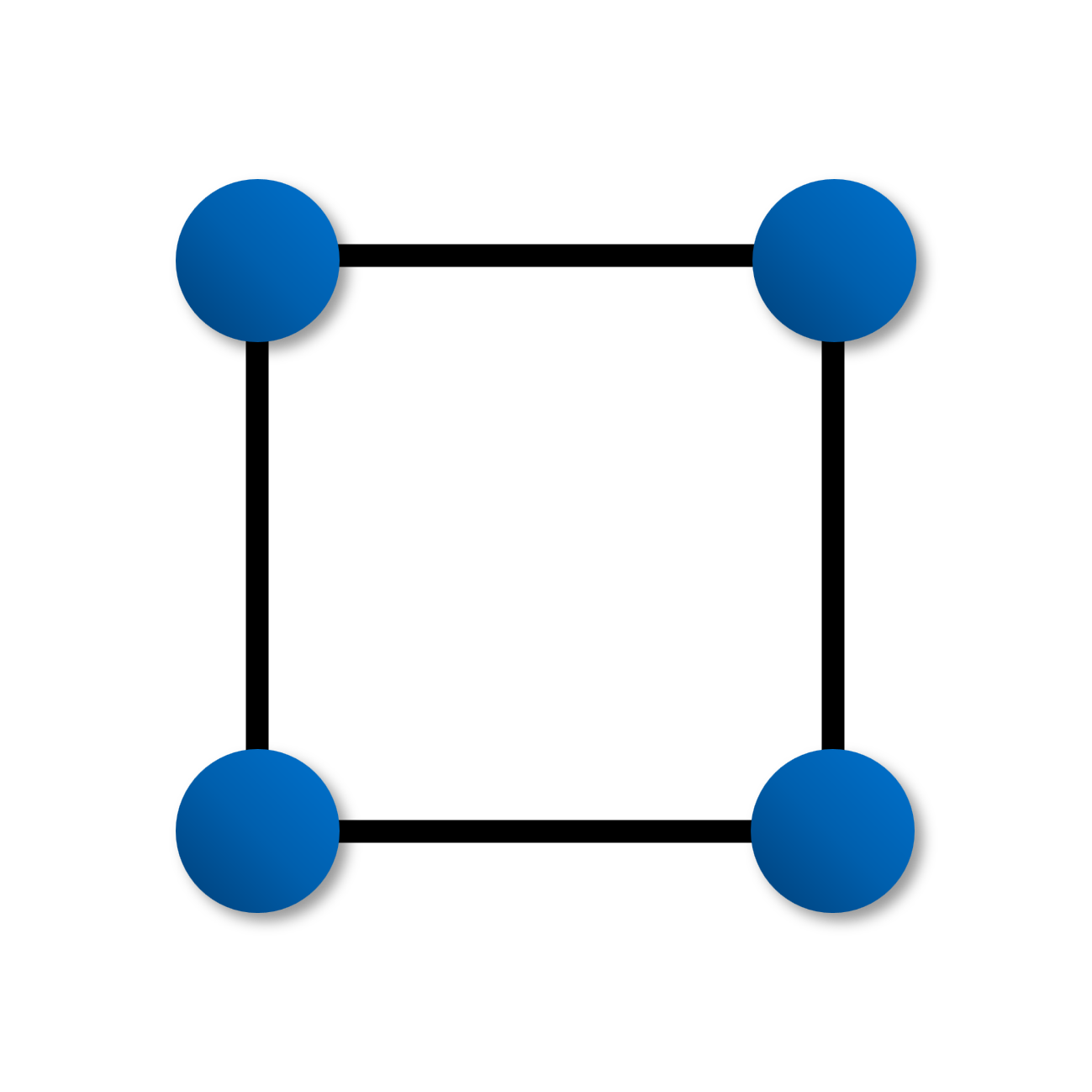}}
& \multirow[c]{2}{*}{1}
& Tl$_4$SnS$_3$
& \multirow[c]{2}{*}{12.68}
& \multirow[c]{2}{*}{CBM}
& \multirow[c]{2}{*}{$\bar{E}_2@1c$} \\
&&&& (P422) &&& \\

&&&
\multirow[c]{2}{*}{2}
& MgO$_2$
& \multirow[c]{2}{*}{12.68}
& \multirow[c]{2}{*}{CBM-2}
& \multirow[c]{2}{*}{$\bar{E}@2e$} \\
&&&& (P422) &&& \\

&&&
\multirow[c]{2}{*}{4}
& CuBr
& \multirow[c]{2}{*}{7.62}
& \multirow[c]{2}{*}{VBM}
& \multirow[c]{2}{*}{$B@4j^*$} \\
&&&& (P4222) &&& \\

\cmidrule(lr){2-8}

& \multirow[c]{2}{*}{2 (checkerboard)}
& \multirow[c]{2}{*}{\icon{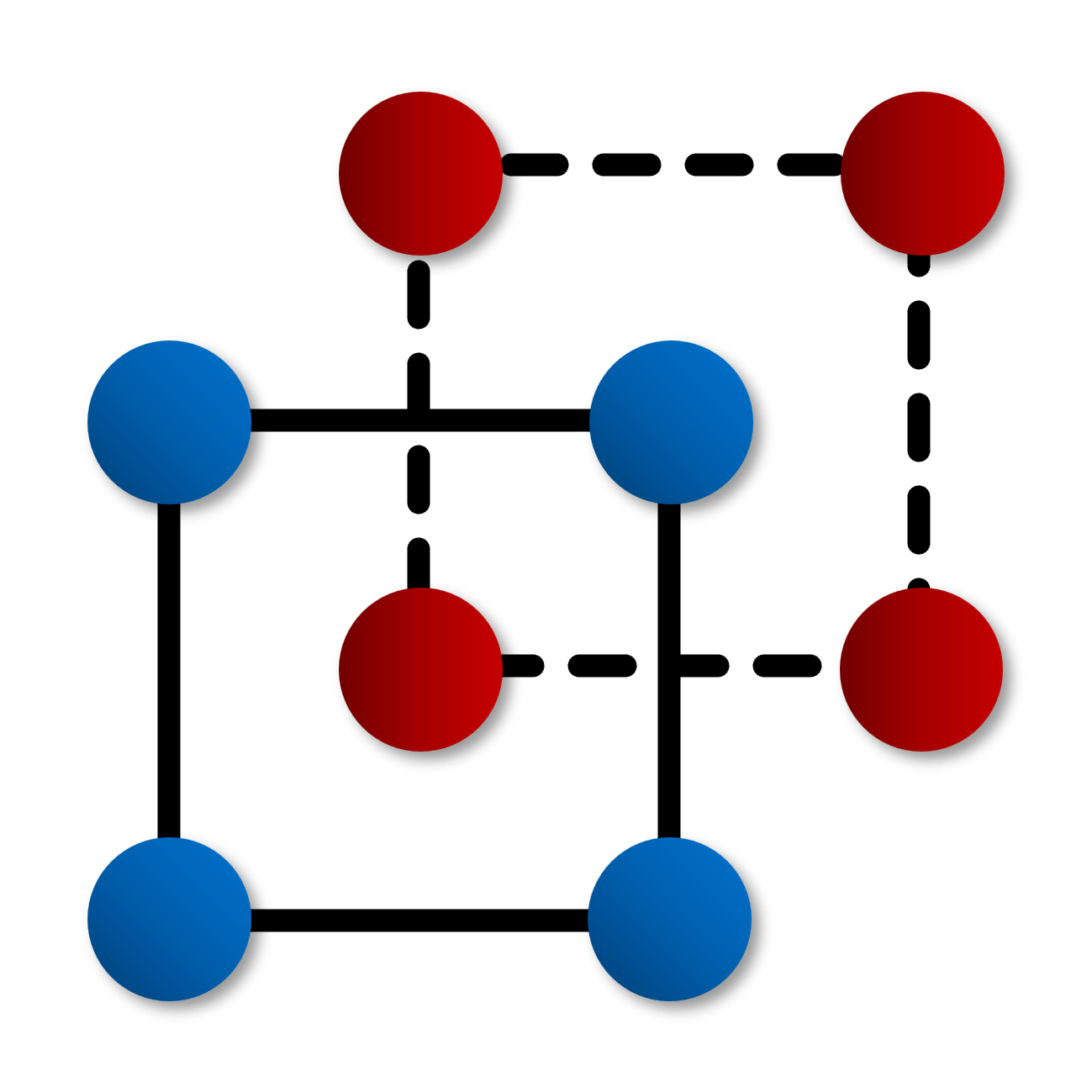}}
& \multirow[c]{2}{*}{1}
& Cu$_2$WS$_4$
& \multirow[c]{2}{*}{8.80}
& \multirow[c]{2}{*}{CBM}
& \multirow[c]{2}{*}{$\prescript{1}{}{\bar{E}}\prescript{2}{}{\bar{E}}@4e$} \\
&&&& (C222) &&& \\

\midrule

\multirow[c]{4}{*}{Rectangular}
& \multirow[c]{2}{*}{1}
& \multirow[c]{2}{*}{\icon{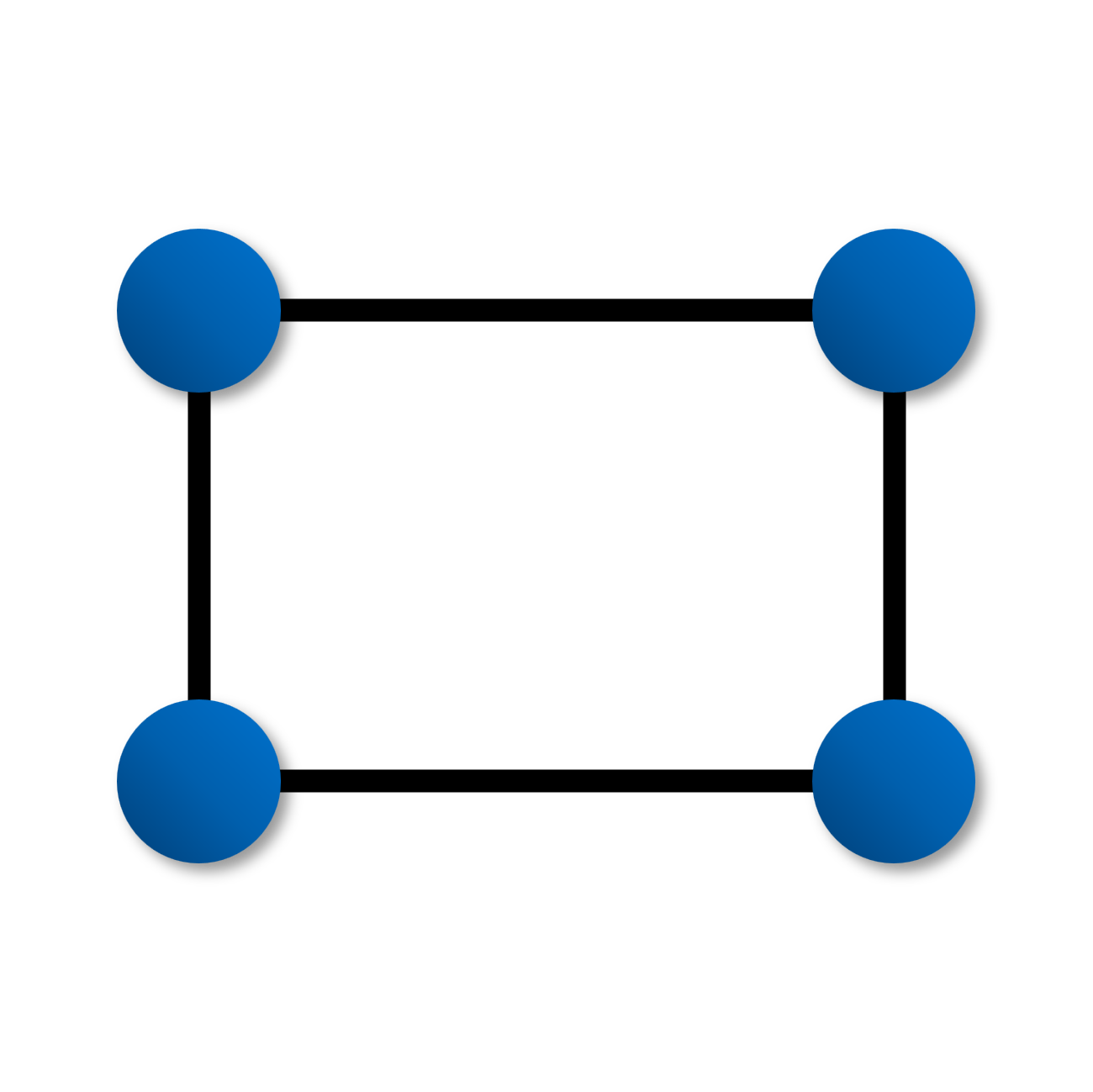}}
& \multirow[c]{2}{*}{1}
& CuHgClSe
& \multirow[c]{2}{*}{17.97}
& \multirow[c]{2}{*}{VBM}
& \multirow[c]{2}{*}{$\prescript{1}{}{\bar{E}}\prescript{2}{}{\bar{E}}@1a$} \\
&&&& (P2) &&& \\

\cmidrule(lr){2-8}

& \multirow[c]{2}{*}{2}
& \multirow[c]{2}{*}{\icon{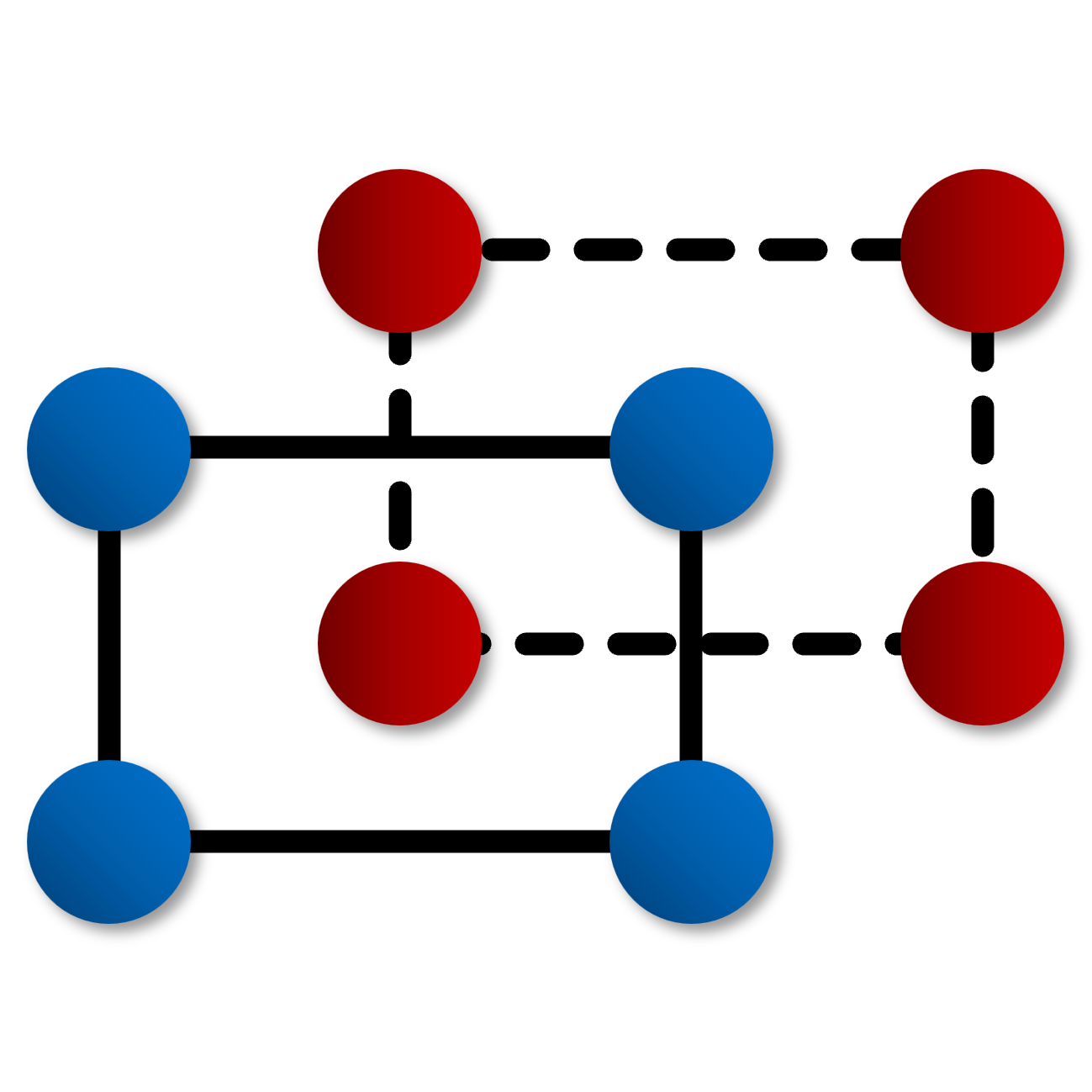}}
& \multirow[c]{2}{*}{1}
& SnS
& \multirow[c]{2}{*}{8.48}
& \multirow[c]{2}{*}{VBM}
& \multirow[c]{2}{*}{$\bar{A}\bar{A}@2a$} \\
&&&& (P2$_1$) &&& \\

\midrule

\multirow[c]{2}{*}{Oblique}
& \multirow[c]{2}{*}{1}
& \multirow[c]{2}{*}{\icon{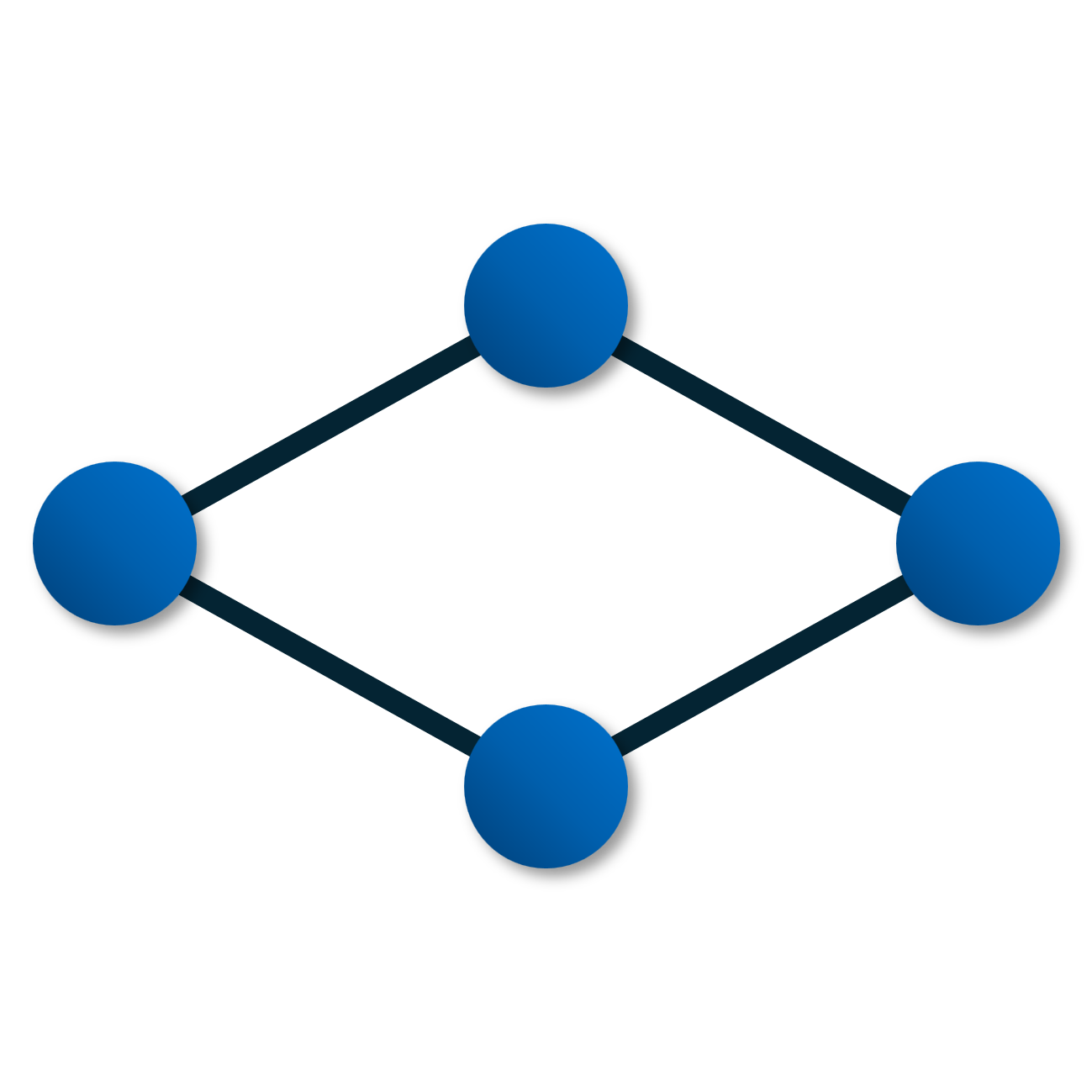}}
& \multirow[c]{2}{*}{2}
& HgPSe$_3$
& \multirow[c]{2}{*}{24.24}
& \multirow[c]{2}{*}{VBM}
& \multirow[c]{2}{*}{$\prescript{1}{}{\bar{E}}\prescript{2}{}{\bar{E}}@2a$} \\
&&&& (C2) &&& \\

\bottomrule
\end{tabular}}
\caption{\fontsize{10pt}{11pt}\selectfont
Representative valley-orbital-symmetry classes of minimal moir\'e lattice models inferred from irreducible-representation (irrep) and elementary band representation (EBR) analysis of isolated band-edge manifolds. The manifolds are grouped by effective lattice type, representative Wyckoff positions and the minimal number of Wannier orbitals $n_w$, with material realizations near the VBM or CBM listed for each class. Here, CBM-2 denotes the second conduction band above the CBM, and an asterisk marks an (E)BR assignment obtained from calculations without SOC. The irrep and EBR labels follow the notation in \textit{Bilbao Crystallographic Server} (BCS)~\cite{aroyo2011crystallography, aroyo2006bilbao1, aroyo2006bilbao2}.}
\label{table:ebrs.rep}
\end{table}

The symmetry-classification branch first resolves the real-space content of an isolated flat-band manifold. Its bandwidth alone does not specify the effective model: one must also determine the number and symmetry of the Wannier centers and the orbital multiplicity at each center. EBR analysis constrains these local degrees of freedom and thereby establishes the Wannier characterization $\mathcal W$.

Table~\ref{table:ebrs.rep} summarizes representative outcomes. Read as a physics table, it separates single-orbital honeycomb, triangular and square Hubbard models from multi-orbital trigonal models, four-orbital square models, checkerboard lattices, kagome-like manifolds and quasi-one-dimensional rectangular systems. The detailed irrep and EBR data are given in SI Section~III; the purpose of the table here is to translate the symmetry information into the low-energy degrees of freedom that enter the effective moir\'e Hamiltonian.

\begin{figure*}[ht]
\includegraphics[width=\textwidth]{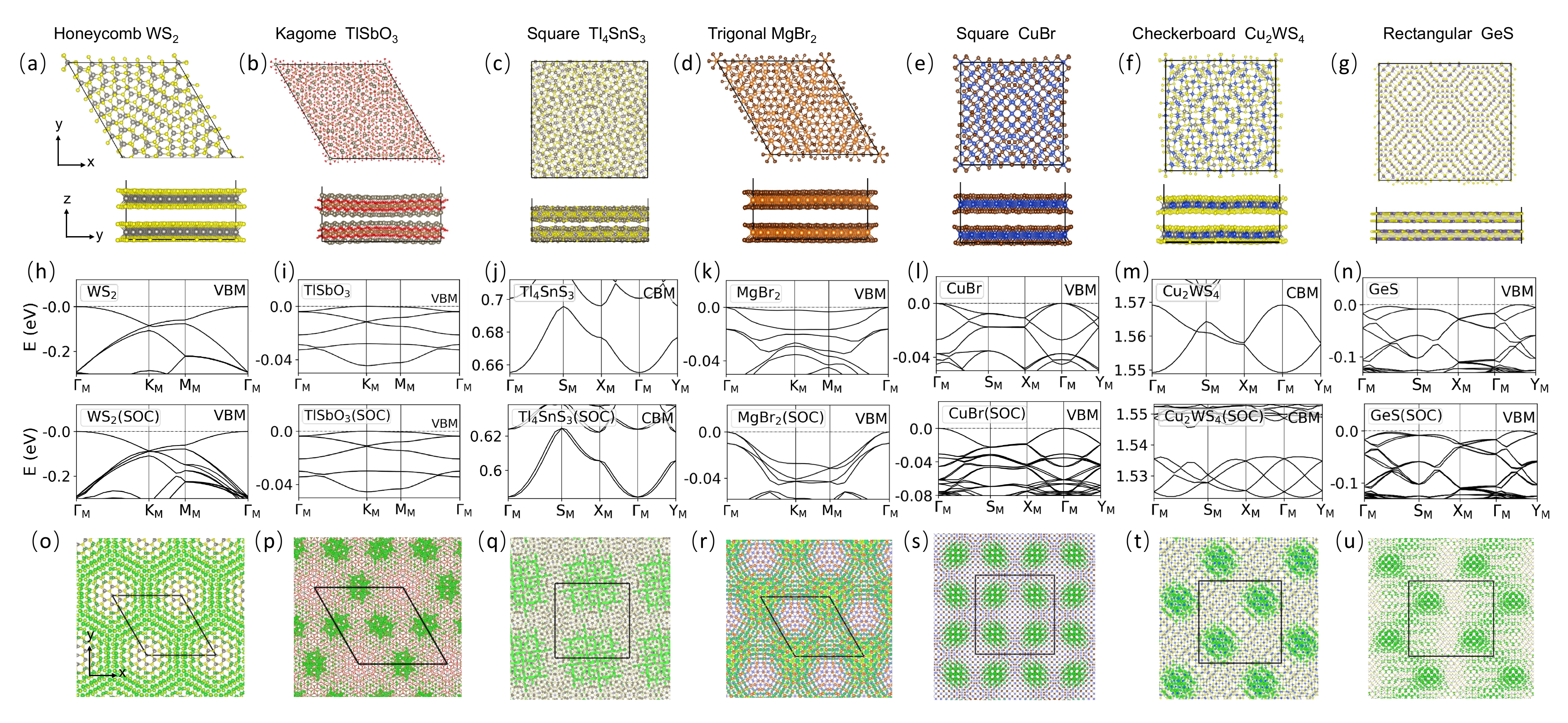}
\centering
\caption{\fontsize{10pt}{11pt}\selectfont  \captiontitle{Flat-band manifolds as emergent Hubbard-model building blocks.}
\textbf{a-g}, Top and side views of relaxed twisted bilayers realizing distinct low-energy Hilbert spaces: honeycomb WS$_2$ at $7.34^\circ$ (\textbf{a}), kagome-like TlSbO$_3$ at $7.34^\circ$ (\textbf{b}), single-orbital square Tl$_4$SnS$_3$ at $12.68^\circ$ (\textbf{c}), multi-orbital trigonal MgBr$_2$ at $7.34^\circ$ (\textbf{d}), four-orbital square CuBr at $7.62^\circ$ (\textbf{e}), checkerboard Cu$_2$WS$_4$ at $8.80^\circ$ (\textbf{f}) and nonsymmorphic rectangular GeS at $9.21^\circ$ (\textbf{g}).
\textbf{h-n}, Corresponding first-principles band structures without spin-orbit coupling (upper panels) and with spin-orbit coupling (lower panels). The relevant flat-band manifolds occur near the valence-band maximum for WS$_2$, TlSbO$_3$, MgBr$_2$, CuBr and GeS, and near the conduction-band minimum for Tl$_4$SnS$_3$ and Cu$_2$WS$_4$.
\textbf{o-u}, Real-space charge-density distributions of the corresponding band-edge flat-band manifolds, shown in green. The black outlines indicate the moir\'e unit cells.}
\label{FigGamma}
\end{figure*}

Single-orbital limits provide the simplest members of this classification. Representative cases include triangular models in twisted PtSe$_2$, honeycomb models in transition-metal dichalcogenides such as WS$_2$, kagome-like manifolds in TlSbO$_3$ and square-lattice models in Tl$_4$SnS$_3$ (Fig.~\ref{FigGamma}). These examples show how the moir\'e potential can localize band-edge states on different emergent real-space lattices while preserving a single-orbital description. In particular, isolated square-lattice flat bands provide a direct route to tunable square-lattice Hubbard physics in a van der Waals setting.

The same analysis exposes richer orbital Hilbert spaces. In MgBr$_2$ (Fig.~\ref{FigGamma}d), the isolated valence-band manifold is described by $p_x$- and $p_y$-like Wannier orbitals on an emergent trigonal moir\'e lattice. This realizes an orbital Hubbard platform in which orbital polarization, Hund-coupled interactions and SOC can all act within the flat-band manifold. CuBr (Fig.~\ref{FigGamma}e) is more unusual: its isolated valence manifold is assigned to $B\uparrow G@4j$, corresponding to four symmetry-related Wannier centers on an emergent square lattice. The low-energy Hilbert space is therefore not a conventional single-band square lattice, but a four-orbital square model selected by moir\'e symmetry.

Multi-site unit cells provide another extension of the single-orbital paradigm. In Cu$_2$WS$_4$ (Fig.~\ref{FigGamma}f), the CBM flat-band manifold is associated with two distinct moir\'e sites, leading naturally to a checkerboard-lattice description. Such manifolds introduce sublattice degrees of freedom and provide realistic settings for frustrated and competing interaction-driven phases. Thus the main physics outcome of this section is a symmetry-based dictionary from first-principles flat bands to the lattice, orbital and sublattice content of candidate Hubbard models.

\section{\textbf{Orbital and nonsymmorphic routes to topological moir\'e bands}}

The same symmetry data also determine the connectivity $\mathcal C$ and topological classification $\mathcal I$ of the target manifold. In twisted transition-metal dichalcogenides, moir\'e reconstruction of spin-valley-locked $K/K'$ states provides a well-established route to topological valence bands \cite{wu2019topological,devakul2021magic,wang2024fractional,nakatsuji2025high}, and interactions within these bands can support fractional Chern physics \cite{park2023observation,zeng2023thermodynamic,cai2023signatures,xu2023observation}. Our calculations recover this conventional setting and reveal two additional routes associated with multi-orbital manifolds and nonsymmorphic band connectivity.

The first route is orbital and spin-orbit driven. In MgBr$_2$, the relevant valence states form a $p_x$-$p_y$-like orbital doublet on an emergent trigonal lattice. Without SOC, these orbitals remain degenerate at $\Gamma_{\mathrm M}$ (Fig.~\ref{FigGamma}k). SOC lifts the degeneracy and yields an isolated four-band manifold separated from neighboring bands. Its irrep content cannot be written as a non-negative integer sum of EBRs, revealing a symmetry-based Wannier obstruction for the isolated manifold. This obstruction originates within the moir\'e flat-band Hilbert space from the interplay of orbital angular momentum, lattice symmetry and SOC, rather than from Berry curvature inherited from a parent $K$ valley.

The same EBR diagnosis identifies additional orbital-driven topological candidates, including twisted CdS, SnSe$_2$ and BiClTe bilayers. These materials differ chemically, but share the same physical feature: isolated multi-orbital moir\'e manifolds whose symmetry representations obstruct a trivial Wannier description. This shows that orbital topology is a general mechanism for moir\'e flat bands, not a special property of a single compound.

The second route is enforced by nonsymmorphic symmetry. Here the central signature is unavoidable band connectivity rather than an isolated Chern-like band. Twisted GeS in space group $P2_1$ (No.~4) provides the representative example (Fig.~\ref{FigGamma}g). With SOC, its flat-band manifold develops degeneracies along high-symmetry lines of the moir\'e Brillouin zone. Compatibility-relation and EBR analyses show that these crossings cannot be removed without breaking the underlying nonsymmorphic symmetry, yielding a symmetry-enforced semimetal within the flat-band manifold. The same mechanism recurs in chemically and crystallographically distinct systems, such as ZrIN($P2_12_12$, No.~18), CuClO$_2$ ($P222_1$, No.~17), and AlP ($P4_122$, No.~91). Their irreps and EBR decompositions are listed in the SI. Together with the conventional $K$-valley setting, the multi-orbital and nonsymmorphic examples establish distinct routes to topological or symmetry-enforced moir\'e band structures.

\section{\textbf{Boundary-valley $\mathbf Q$-lattices and quasi-one-dimensional flat bands}}

\begin{figure}[ht]
\includegraphics[width=0.5\columnwidth]{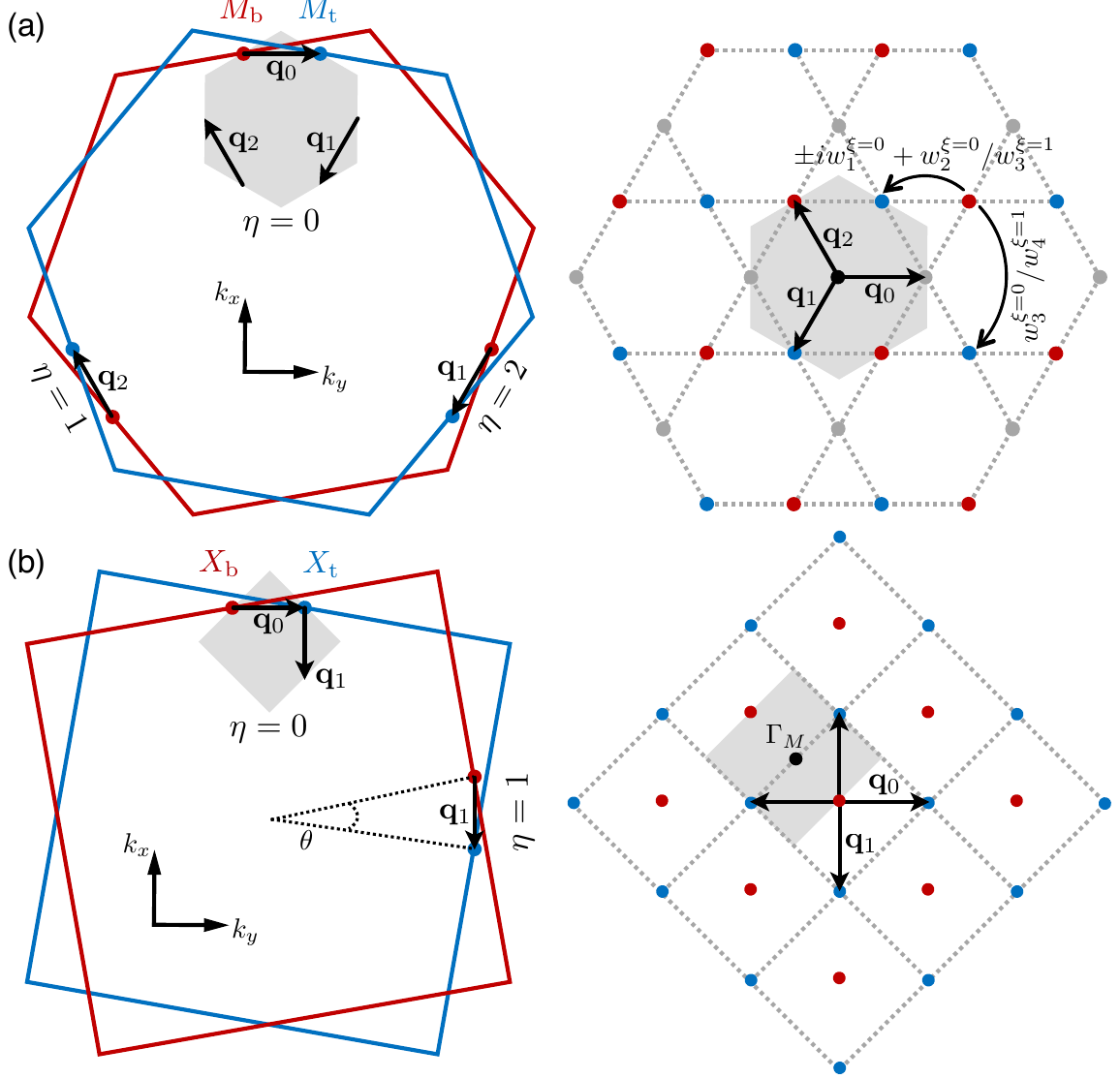}
\centering
\caption{\fontsize{10pt}{11pt}\selectfont \captiontitle{Momentum-space geometry of $M$- and $X$-valley moir\'e systems.} Geometric relationship between the monolayer and moir\'e Brillouin zones for (a) $M$-point and (b) $X$-point moir\'e materials. The right panels show the corresponding reciprocal-space $\mathbf Q$-lattices in valley $\eta=0$ and representative moir\'e coupling terms $T_{\mathbf Q,\mathbf Q'}$.}
\label{fig:m-point-continuum}
\end{figure}

\begin{figure*}[ht]
\centering
\includegraphics[width=\textwidth]{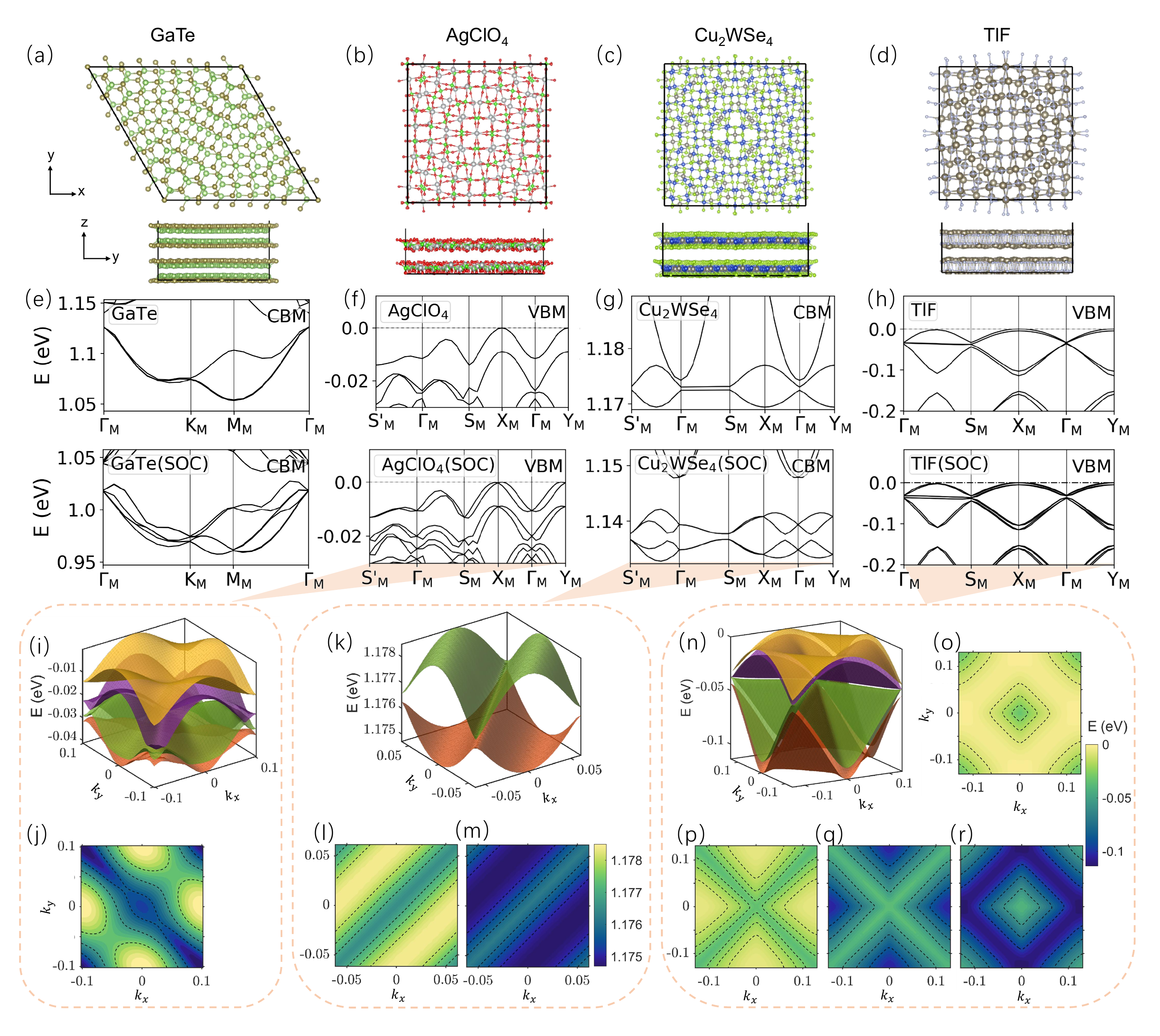}
\caption{ \captiontitle{Boundary-valley moir\'e bands and quasi-one-dimensional dispersions.} (a-d) Top and side views of twisted bilayer GaTe at 7.34$^{\circ}$ (a), AgClO$_4$ at 12.68$^{\circ}$ (b), Cu$_2$WSe$_4$ at 8.80$^{\circ}$ (c) and TlF at 12.68$^{\circ}$ (d). (e-h) Corresponding electronic band structures without spin-orbit coupling (upper panels) and with spin-orbit coupling (lower panels). (i) Three-dimensional band dispersion of AgClO$_4$ at 12.68$^{\circ}$ near the VBM; the valence-band edge is highlighted in yellow and its two-dimensional contour map is shown in (j). (k) Three-dimensional dispersion of the bottom two conduction bands of Cu$_2$WSe$_4$ at 8.80$^{\circ}$ with contour maps in (l,m). (n) Three-dimensional dispersion of the top four valence bands of TlF at 12.68$^{\circ}$ with contour maps in (o-r).}
\label{FigMX}
\end{figure*}

The parent-valley momentum $\mathbf{k}_0$ also controls the reciprocal-space structure of the moir\'e problem. Boundary valleys provide the clearest realization of the continuum-model branch because their symmetry-related copies form extended networks of plane-wave states connected by moir\'e reciprocal vectors. $k_0$ determines the geometry of the $\mathbf Q$-lattice, whereas the parent-band orbital content and $G_{\mathrm M}$ specify the internal basis and allowed couplings. Recently, hexagonal materials with band extrema at the $M$ points were shown to realize such physics \cite{cualuguaru2025moire,lei2025moire}. Unlike $\Gamma$- or $K$-valley moir\'e materials, the three time-reversal-preserving $M$ valleys are related by threefold rotation, $C_{3z}M^{\eta}=M^{(\eta+1)\bmod 3}$, and generate a kagome-like $\mathbf Q$-lattice in momentum space (Fig.~\ref{fig:m-point-continuum}a).

For valley $\eta=0$, a minimal continuum description of the $M$-point moir\'e bands is
\begin{align}
[h_{Q,Q'}(k)]_{ls;l's'} =& \delta_{Q,Q'}\delta_{ll'}\delta_{ss'} \bigg[ \frac{(k_x - Q_x)^2}{2m_x} + \frac{(k_y - Q_y)^2}{2m_y} \bigg] \notag\\&+ [T_{Q,Q'}]_{ls;l's'},
\label{eq:m-point-continuum}
\end{align}
where $m_x$ and $m_y$ are the anisotropic effective masses of the parent valley, $l$ and $s$ label layer and spin, and $\mathbf Q$ denotes momenta in the emergent reciprocal-space lattice. To lowest order, the tunneling matrix contains interlayer terms such as $[T_{Q,Q'}^{\xi=0}]_{ls;(-l)s}=(\pm i w_1^{\xi=0}+w_2^{\xi=0})\delta_{Q\pm q_0,Q'}+w_3^{\prime AA}\delta_{Q\pm(q_1-q_2),Q'}$, where $q_{0\ldots2}$ are scattering vectors connecting plane-wave states from the two layers. The index $\xi=0,1$ distinguishes the two symmetry settings $P312$ and $P321$.

The key physical effect is a momentum-space nonsymmorphic symmetry, generated in the rigid-shift limit by an effective mirror symmetry $\tilde M_z$. This emergent momentum-space symmetry is distinct from the crystallographic nonsymmorphic symmetry responsible for the semimetallic connectivity discussed above. It enforces energetic nesting between $\mathbf k$ and $\mathbf k+\mathbf q_\eta$ and produces quasi-one-dimensional dispersions that are distinct from the trivial anisotropy associated with a large monolayer mass ratio. Earlier work established this mechanism in 1T-SnSe$_2$ and 1T-ZrS$_2$ \cite{cualuguaru2025moire} and analysed the associated interacting six-flavor Hubbard models \cite{li2025emergentinteractingphasesstrong}.

The present first-principles atlas identifies additional $M$-point systems, including AA-stacked GaTe, HfS(Se)$_2$, Al$_2$MgS$_4$ and AgGaP$_2$Se$_6$, several of which have appreciable SOC. For example, AA-stacked GaTe has moir\'e symmetry $P321$ and parent space group $P\bar{6}m2$ (No.~187) (Fig.~\ref{FigMX}a). In the zero-twist-angle limit, an effective inversion symmetry inherited from the horizontal mirror of the untwisted bilayer acts nonsymmorphically in momentum space. The simplified continuum model in Eq.~\eqref{eq:m-point-continuum} reproduces the DFT band structure with fitted spin-orbit terms, as detailed in SI Section~V.E.

The same boundary-valley principle extends beyond hexagonal $M$ points. In square lattices with low-energy states at $X$, or rectangular lattices with states at $X/Y$, twisting produces a checkerboard $\mathbf Q$-lattice (Fig.~\ref{fig:m-point-continuum}b) that is the square-lattice analogue of the $M$-point kagome-like network. Representative examples are AgClO$_4$, Cu$_2$WSe$_4$ and TlF (Fig.~\ref{FigMX}b-d). Thus the decisive ingredient is the location and symmetry-related multiplicity of the parent boundary valley, rather than the chemistry of a particular material family.

Cu$_2$WSe$_4$ illustrates the resulting physics. The monolayer crystallizes in space group $P\bar{4}2m$ (No.~111), while the twisted bilayer has space group $C222$ (Fig.~\ref{FigMX}c). At $\theta=8.8^\circ$, the low-energy conduction bands remain nearly degenerate around $\Gamma$, $S$ and $S'$, and develop an almost dispersionless near-nodal feature along $\Gamma$-$S$; the perpendicular $\Gamma$-$S'$ direction is substantially more dispersive and separated by a sizable gap. The three-dimensional dispersion and constant-energy contours in Fig.~\ref{FigMX}k-m make this quasi-one-dimensional character explicit. Boundary-valley twisting therefore provides a general route to quasi-one-dimensional flat bands in both hexagonal and non-hexagonal moir\'e materials. A more complete continuum treatment of the square-lattice $X$-point problem is being developed in a follow-up work \cite{twistedX}; here we focus on its first-principles manifestations and on how the underlying valley geometry organizes the quasi-one-dimensional dispersions shown in Fig.~\ref{FigMX}.

\section{\textbf{Generic-momentum valleys and symmetry-selected multi-valley multiplets}}

\begin{figure*}[ht]
\centering
\includegraphics[width=\textwidth]{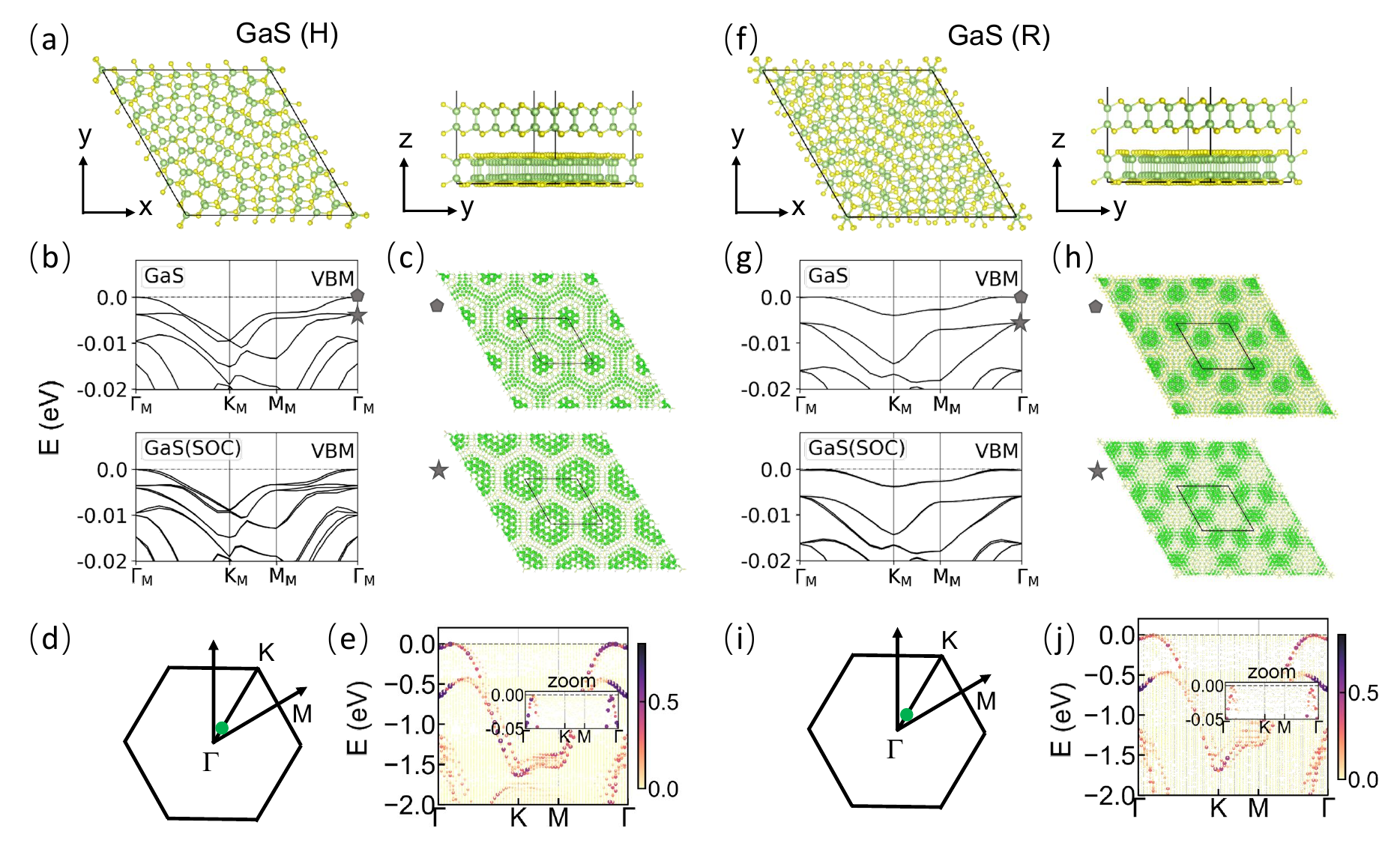}
\caption{\fontsize{10pt}{11pt}\selectfont \textbf{Symmetry-selected multiplets from nearby generic-momentum valleys in twisted GaS.}
\textbf{a-e}, Results for twisted bilayer GaS derived from the H phase; \textbf{f-j}, corresponding results for the R phase.
\textbf{a,f}, Top and side views of the fully relaxed moir\'e structures at a twist angle of $7.34^\circ$. The black outlines indicate the moir\'e unit cells.
\textbf{b,g}, First-principles moir\'e band structures calculated without spin-orbit coupling (upper panels) and with spin-orbit coupling (lower panels). The pentagons and stars mark the valence-band manifolds whose real-space distributions are shown in \textbf{c} and \textbf{h}.
\textbf{c,h}, Charge-density distributions of the selected flat-band manifolds, shown in green.
\textbf{d,i}, Monolayer Brillouin zones showing the non-high-symmetry-point valleys near $\Gamma$ that contribute to the moir\'e valence-band states; the green markers denote representative valley positions and the symmetry-related valleys are generated by the threefold rotation.
\textbf{e,j}, Moir\'e band structures unfolded onto the primitive monolayer Brillouin zone for the H- and R-phase structures, respectively. The colour scale denotes the unfolding spectral weight, and the insets magnify the states near the valence-band maximum. In the H-phase structure, the nearby valleys hybridize into a connected three-band manifold, whereas in the R-phase structure they are reorganized into separate one- and two-band sectors.}
\label{FignHS}
\end{figure*}

High-symmetry valleys provide a simple starting point for moir\'e-band classification because different valley sectors are often well separated in momentum space. In familiar $K$-valley systems, the two valleys can often be treated as approximately independent flavors. A related separation occurs for different $M$ valleys in hexagonal systems and $X/Y$ valleys in square or rectangular lattices, even though the boundary-valley $\mathbf Q$-lattices discussed above introduce additional structure.

Generic-momentum valleys, referred to below as nHSP valleys, provide a different limit. Because their positions are not fixed by high-symmetry momenta, the distances among symmetry-related valleys and their moir\'e-induced hybridization are material dependent. When several such valleys lie close to one another near the band edge, the moir\'e potential can mix them into a single low-energy manifold rather than leaving them as independent valley flavors.

Twisted GaS provides a representative example, with similar behavior in GaSe and InSe (Fig.~\ref{FignHS}). In the H-phase-derived structure with moir\'e space group $G_\mathrm{M}=P321$, band unfolding shows that the valence-band states originate from three $C_3$-related nHSP valleys near $\Gamma$ in the monolayer Brillouin zone (Fig.~\ref{FignHS}d,e). These valleys hybridize into an isolated connected three-band manifold near the moir\'e valence-band maximum (Fig.~\ref{FignHS}b), which is reminiscent of a kagome manifold: one pair of bands is degenerate at $\Gamma_{\mathrm M}$, whereas a different pair is degenerate at $K_{\mathrm M}$. The EBR decomposition $A_2\uparrow G@1a\oplus E\uparrow G@1a$ identifies the manifold as a local $1+2$ orbital multiplet at the same moir\'e Wyckoff position. It is therefore not a kagome representation induced from one Wannier orbital on three symmetry-related sites; it is a local $A_2\oplus E$ multiplet formed from coupled nearby valleys and organized by the $P321$ moir\'e symmetry.

This outcome is not determined by valley proximity alone. In the R-phase-derived GaS structure, the parent monolayer has $P\bar{3}m1$ symmetry, whereas the corresponding moiré structure belongs to $P312$ rather than $P321$ and no longer contains the same local stacking geometry (Fig.~\ref{FignHS}f). In the spinless description, the top valence bands are resolved into separate one- and two-band sectors instead of the single connected $1+2$ manifold (Fig.~\ref{FignHS}g). The second band set realizes the trigonal $p_x\pm i p_y$ model listed in Table~\ref{table:ebrs.rep}. The H/R comparison shows that the effective Hamiltonian is selected jointly by $k_0$ and $G_\mathrm{M}$: nearby nHSP valleys supply the ingredients, but symmetry determines how those ingredients are reorganized into low-energy multiplets.

The resulting flat bands are qualitatively different from conventional high-symmetry-valley systems. Instead of weakly coupled valley flavors, the GaS family realizes valley-derived orbital multiplets whose internal structure emerges from moir\'e hybridization. This provides a route to correlated multi-orbital moir\'e Hubbard physics in which the active degrees of freedom are moir\'e-localized orbital multiplets formed from several nearby generic-momentum valleys, rather than approximately independent valley flavors.

\section{\textbf{Discussion}}

In conclusion, the atlas converts first-principles band-edge manifolds into a classification of effective moir\'e models. EBR and compatibility analyses identify their real-space lattice and Wannier content, symmetry-based Wannier obstruction and enforced band connectivity. Complementarily, the $\mathbf Q$-lattice analysis reveals the momentum-space structures associated with boundary valleys and clarifies the origin of quasi-one-dimensional dispersions in selected systems. Together, these real- and momentum-space descriptions organize the more than 600 relaxed commensurate bilayers into a set of physical mechanisms rather than a list of candidate materials.

Five motifs emerge. First, orbital and Wyckoff-position engineering produces canonical and non-canonical Hubbard models, including single-orbital honeycomb and square models, multi-orbital trigonal and four-orbital square models, and multi-site checkerboard or kagome-like manifolds. Second, topological flat bands can arise from orbital multiplets with SOC, complementing the conventional $K$-valley setting. Third, nonsymmorphic moir\'e symmetries can enforce semimetallic connectivity directly within flat-band manifolds. Fourth, the geometry of symmetry-related boundary valleys controls the reciprocal-space $\mathbf Q$-lattices underlying the quasi-one-dimensional bands found in $M$- and $X$-valley systems. Fifth, coupled multi-valley manifolds with kagome-like connectivity occur in several systems whose parent band edges lie at nHSPs.

This physics-centered classification provides concrete targets for future experimental and theoretical work. The material examples identified here specify the lattice geometry, orbital content, valley origin and symmetry constraints of the low-energy bands, which are the inputs needed for interaction estimates, continuum-model refinement and experimental probes by gating, transport, optical spectroscopy or scanning tunneling microscopy. More broadly, the valley-orbital-symmetry map offers a route for designing correlated, topological and symmetry-enforced moir\'e phases from the electronic structure of the parent two-dimensional crystal.

\begin{acknowledgments}
We acknowledge helpful discussions with Kin Fai Mak and Jie Shan. L.X., E.W., T.Z., Y.G. and Q.X. acknowledge the support by the National Key Research and Development Program of China (Grant No. 2022YFA1403501), the National Natural Science Foundation of China (Grants No. 12474174 and No. 12504218), Hangzhou Tsientang Education Foundation and the Max Planck Partner group programme. We acknowledge computational support from the Max Planck Computing and Data Facility, the Computing Center of Tsientang Institute for Advanced Study, and the Hefei Advanced Computing Center. D.M.K. acknowledges funding by the Deutsche Forschungsgemeinschaft
(DFG, German Research Foundation) within the Priority Program SPP 2244 “2DMP”—443274199 and under Germany’s Excellence Strategy—Cluster of Excellence Matter and Light for Quantum Computing (ML4Q) EXC 2004/1–390534769. A.F. is funded by the Deutsche Forschungsgemeinschaft (DFG, German Research Foundation) - 572935092.  We acknowledge computational resources provided by RWTH Aachen University under Project No. rwth0763. This work was supported by the Max Planck-New York City Center for Nonequilibrium Quantum Phenomena. This work was supported by the European Research Council (ERC Synergy Grant Agreement No. 101167294 (UnMySt)), the Cluster of Excellence “CUI: Advanced Imaging of Matter” of the Deutsche Forschungsgemeinschaft (DFG)—EXC 2056—Project ID 390715994 and Grupos Consolidados y Alto Rendimiento UPV/EHU, Gobierno Vasco (IT1453-22). The Flatiron Institute is a division of the Simons Foundation.
Y.J. was supported by the European Research Council (ERC) under the European Union’s Horizon 2020 research and innovation program (Grant Agreement No. 101020833), as well as by the IKUR Strategy under the collaboration agreement between Ikerbasque Foundation and DIPC on behalf of the Department of Education of the Basque Government. 
H.P. was supported by the Ministry for Digital Transformation and of Civil Service of the Spanish Government through the QUANTUM ENIA project call - Quantum Spain project, and by the European Union through the Recovery, Transformation and Resilience Plan - NextGenerationEU within the framework of the Digital Spain 2026 Agenda. 
B.A.B. is supported by the Gordon and Betty Moore Foundation through Grant No. GBMF8685 towards the Princeton theory program, the Gordon and Betty Moore Foundation’s EPiQS Initiative (Grant No. GBMF11070), the Global Collaborative Network Grant at Princeton University, the Simons Investigator Grant No. 404513, the NSF-MERSEC (Grant No. MERSEC DMR 2011750), the Simons Collaboration on New Frontiers in Superconductivity (Grant No. SFI-MPS-NFS-00006741-01), Princeton Catalysis Initiative (PCI), the Schmidt Foundation at the Princeton University and the National Science Foundation through the AI Research Institutes program Award No. DMR-2433348.
\end{acknowledgments}

\section*{Data availability}

The structural, electronic-structure and symmetry-analysis data
supporting the findings of this study are available through the
Twisted Bilayer Moiré Superlattice Database (TBMSD) at
\url{https://2dstack.tias.ac.cn/tbmsd/}. Additional data are
available from the corresponding authors upon reasonable request.

\bibliography{ref}
\end{document}